\documentclass[aps,prb,reprint,groupedaddress,nofootinbib,showkeys]{revtex4-2}

\usepackage{graphicx}
\usepackage{dcolumn}
\usepackage{bm}
\usepackage{amsmath}
\usepackage{newunicodechar}
\newunicodechar{∕}{\textfractionsolidus}

\renewcommand{\thefootnote}{\fnsymbol{footnote}}

\begin{document}
\raggedbottom

\title{\textbf{Thomas-Fermi screening of electrostatic fields in a type-I superconductor}}

\author{Nikhil Seeja Sivakumar}
\author{Tyler James}
\author{Frederick Carlisle}
\author{Philip Moriarty} 
\author{Brian Kiraly} \email{brian.kiraly@nottingham.ac.uk}
\affiliation{School of Physics \& Astronomy, University of Nottingham, Nottingham NG7 2RD, UK}

\date{\today}

\begin{abstract}
The empirical London equations make distinct predictions for a superconductor's longitudinal and transverse electrodynamic response. Conventionally, this framework attributes the Meissner effect to the transverse component, with the longitudinal response predicted to remain unchanged on entering the superconducting state, i.e. a static electric field is screened over the Thomas-Fermi length ($\lambda_{\mathrm{TF}} \sim 1$~\AA), just as in the nonsupercondcuting metal. J.~E.~Hirsch~[\textit{Phys.~ Rev.~B}~\textbf{69},~214515~(2004)] has developed an alternative formalism, which predicts that the longitudinal screening length should instead be governed by the London penetration depth as the system is cooled below the superconducting transition temperature ($T_{\mathrm{c}}$). Here we use a combination of qPlus atomic force microscopy and scanning tunneling microscopy/spectroscopy of single-crystal Pb(111) at $\sim$~340~mK~ \mbox{($T/T_{\mathrm{c}}\sim0.05$)} in an attempt to detect any modifications in electrostatic screening of the longitudinal tip field upon entering the superconducting phase. We quench superconductivity using a magnetic field of 200 mT normal to the Pb(111) surface. By measuring force-distance curves and local field-emission resonance spectra at the same sample position in the presence and absence of the magnetic field, we  constrain the change in screening between the normal and superconducting state to less than $\sim 1 \%$. This is between two to three orders of magnitude smaller than that predicted by Hirsch's theory, and entirely consistent with unmodified Thomas-Fermi screening across the superconducting transition.
\end{abstract}

\maketitle
The electrodynamic response of a superconductor (SC) has been debated since the  London formulation was first put forward almost a century ago~\cite{London1935a}. In traditional BCS theory, the SC order parameter has both an amplitude and phase, and the latter must be treated in concert with the vector potential to realize a fully gauge-invariant formulation of SC electrodynamics~\cite{Anderson1958, Nambu1960,Koyama2004}. The conventional approach attributes magnetic field screening to the transverse electromagnetic response over length scales of the London penetration depth ($\lambda_{\mathrm{L}})$ — thereby producing the Meissner effect — whereas the static longitudinal response screens electric fields over the much shorter Thomas-Fermi screening length ($\lambda_{\mathrm{TF}}$). J. E. Hirsch has instead argued, through a series of occasionally provocative papers~\cite{Hirsch2003, Hirsch2004, Hirsch2012, Hirsch2015, Hirsch2020b, HirschReply, hirsch2020a}, that the invariance requirement is more naturally satisfied in the Lorenz gauge, under which the SC wavefunction is equally rigid against electric and magnetic perturbations, so that the natural screening length for \textit{both} \textbf{E} and \textbf{B} fields is $\lambda_{\mathrm{L}}$. Consequently, the static electrostatic screening length constitutes a direct experimental discriminant between these competing descriptions of the SC state. Beyond its significance for the foundations of SC, resolving this question is of particular relevance for emerging SC technologies, including electrostatically controlled SC devices~\cite{Antipov2018, Chen2017}, SC field-effect transistors~\cite{DeSimoni2018, Paolucci2021}, and atomically thin SC~\cite{Ye2012, Saito2015, Tsen2016}, where the interaction between electric fields and the condensate is of central importance.

Most experimental studies infer screening properties non-locally in capacitive geometries~\cite{London1936,Glover1960} or indirectly through measurements of microwave surface impedance~\cite{Mattis1958, Halbritter1971, Turneaure1968, Junginger2017}, or optical/terahertz conductivity ~\cite{Glover1957,Pracht2013,Sim2017} rather than through direct probes of static electric-field screening. In contrast, a local probe is uniquely capable of measuring the spatial dependence of the electrostatic interaction itself, providing direct access to the screening response of the condensate and disentangling screening from alternative coexisting surface signals. Recognizing this, Hirsch proposed a scanning probe experiment designed to detect the predicted modification of electrostatic screening in the SC state~\cite{Hirsch2015}. Peronio and Giessibl~\cite{Peronio2016} elegantly adapted that proposal to a qPlus atomic force microscope (AFM) to measure the electrostatic interaction between the Smoluchowski dipole~\cite{Smoluchowski1941} at the probe apex and a Nb(110) surface, searching for signatures of the predicted change in screening length. While their experiments established an important benchmark, the achievable force sensitivity prevented a definitive determination of the screening response, leaving the question unresolved.

Here, we overcome these limitations by combining the force sensitivity of a cryogenic qPlus sensor with the local spectroscopic capabilities of the scanning tunneling microscope (via field emission resonances~\cite{Becker1985,Binnig1985, Ploigt2007, Lin2007, Aladyshkin2020}), enabling a direct real-space measurement of the electrostatic screening response of the SC condensate. To provide the clearest possible discrimination, we operate at approximately 340 mK, where $T/T_{\mathrm{c}} \sim 0.05$ and $\lambda_L \sim 53$~nm, and we quench SC with a magnetic field (Fig. 1) rather than a temperature ramp~\cite{Peronio2016}. The combined effect is to dramatically reduce measurement uncertainty while simultaneously enhancing the magnitude of the effective observable: Hirsch's prediction corresponds to three orders of magnitude difference in the screening ($\lambda_L /\lambda_{TF} \sim 10^3$) for bulk Pb~\cite{Muino2011}.

We focus on screening of the Smoluchowski dipole at the apex of the scanning probe~\cite{Peronio2016} by the underlying Pb(111) surface (sketched schematically in Fig.~1(a),~(b)). For our purposes, a perfect metal is represented by the presence of an image dipole whose magnitude and position are exactly equal and opposite that of the tip (Fig.~1(b)). The presence of imperfect screening (Fig. 1(a)) instead causes the voltage to be dropped over a volume inside the sample -- as for semiconductors where the Fermi level can respond to the tip-induced field~\cite{Weimer1989} -- reducing the magnitude of the effective image dipole at the tunnel junction. 

Following sputter-annealing, an atomically flat Pb(111) surface was obtained (Fig.~1(c)). (See the Supplementary Material for detailed information on sample preparation and measurement protocols.) In the absence of a magnetic field, the SC gap of the Pb(111) surface is well-resolved in $\mathrm{d}I/\mathrm{d}V$ spectra (Fig.~1(d), spectrum in blue). Throughout our experiments, care was taken to ensure that the differential conductance spectra were also free of any indication of an additional SC gap that might arise from appreciable quantities of lead adsorbed on the tip. Application of a magnetic field of 200 mT normal to the sample surface quenches the sample SC, as evidenced by the removal of the characteristic gap around the Fermi level in the $\mathrm{d}I/\mathrm{d}V$ spectrum (Fig.~1(d), spectrum in gray). The presence of fiducial surface features such as surface defects and step edges ensure that the tip can be reliably and reproducibly re-positioned.

\begin{figure}[t!]
\includegraphics[width=1 \linewidth]{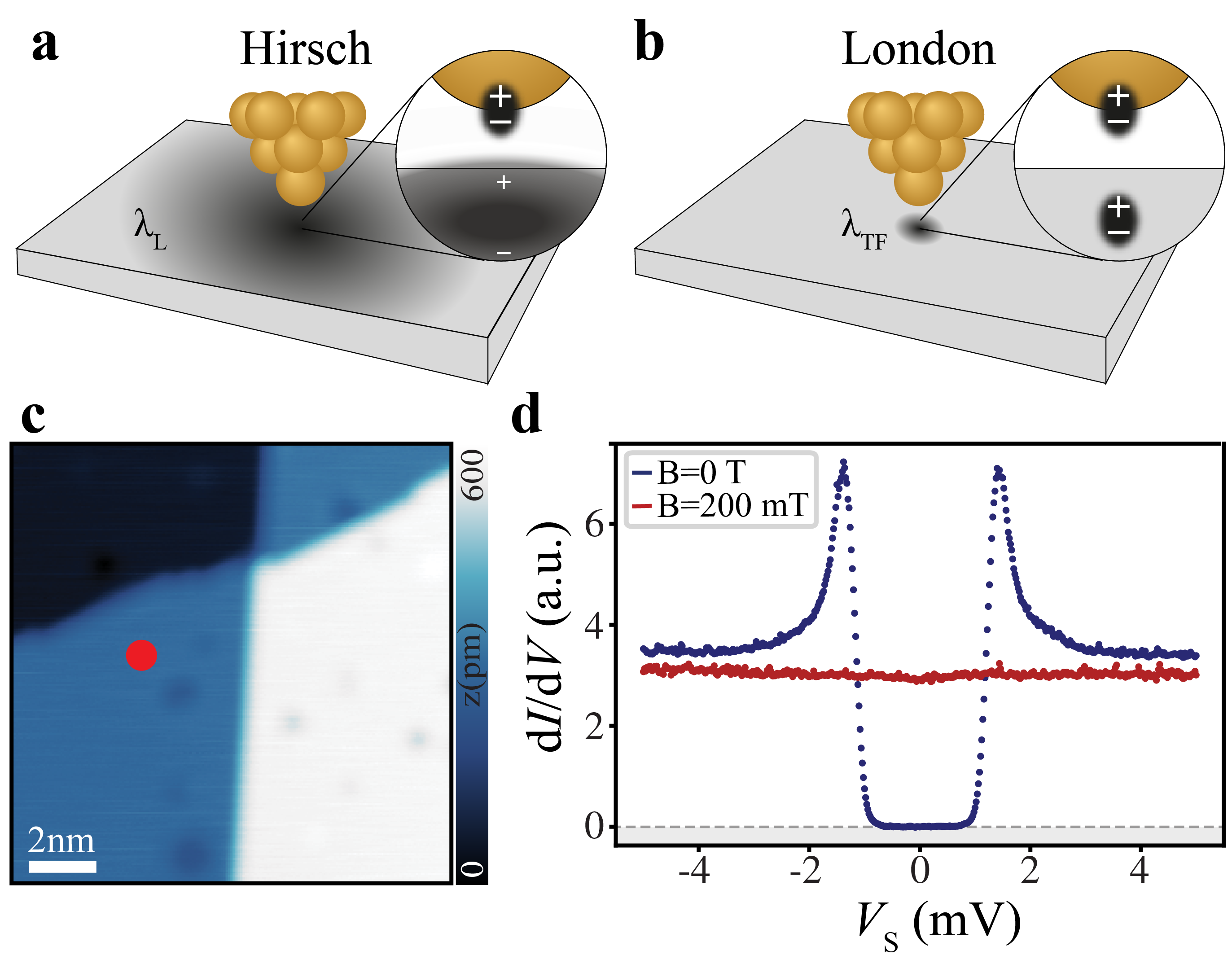}
\caption{Sketch showing the screening of the tip dipole by the superconductor in the case of (a)~Hirsch's alternative theory~\cite{Hirsch2003} (screening length $\sim \lambda_{\mathrm{L}}$, the London penetration depth) and (b)~traditional London (BCS) theory (Thomas-Fermi screening length, $\lambda_{\mathrm{TF}}$). (c)~STM image of Pb(111) surface ($V_{\mathrm{S}}:100$~mV, $I_{\mathrm{t}}$:~100~pA). (d)~d$I$/d$V$ spectra measured in the presence (grey) and absence (blue) of a 200~mT magnetic field applied normal to the Pb(111) sample surface. Stabilisation parameters: $V_{\mathrm{S}}: 5$~mV, $I_{\mathrm{t}}$: 100~pA, Lock-in amplifier modulation amplitude: 20~$\mu$V.}. 

\label{Fig1}
\end{figure}

\begin{figure}[b!]
\includegraphics[width=1\linewidth]{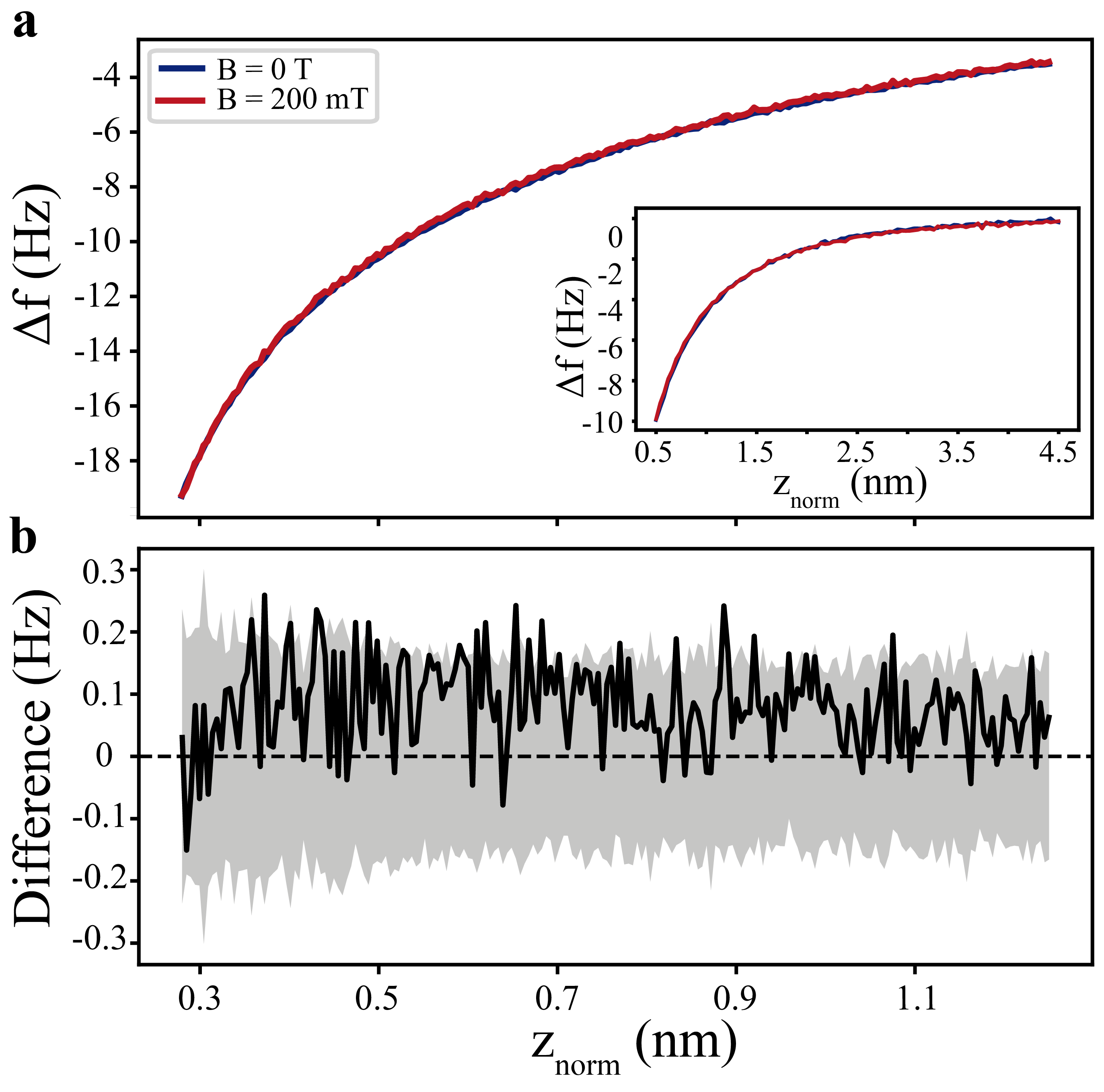}
\caption{(a) Averaged set of $\Delta f(z)$ measurements on the Pb(111) surface with $B=0$~T~(blue; $N=60$ curves) and $B=200$~mT~(red; $N=24$), respectively. (Inset) Similar measurement over a much larger $z$ range. (b) Difference signal between $B=0$~T and $B=200$~mT, with the shaded region given by the 99\% confidence interval associated with the pooled standard error. Stabilisation parameters: $V_{\mathrm{S}} = 5$~mV, $I_{\mathrm{t}}$ = 100~pA, oscillation amplitude = 100 pm.}. 
\label{Fig2}
\end{figure}

In the first set of experiments, we followed the methodology introduced by Peronio and Giessibl\cite{Peronio2016}: frequency-shift-vs-tip-sample-separation ($\Delta f(z)$) curves measured at the same position on the Pb(111) surface are used to determine whether there are changes in the force gradient experienced by the qPlus tuning fork sensor in the superconducting vs normal states. Averaged $\Delta f(z)$ curves acquired at the position denoted by the red dot in Fig.~1(c) are shown in Fig.~2(a). All sets of measurements were made at the same lateral position (to a precision of better than 1\AA) and with identical initial feedback/stabilization parameters far from the superconducting gap onset to set the starting tip height (i.e. the initial $z$ value). The $z$ data is then normalized by setting $z=0$ when $G = G_0$ upon extrapolation of $I-z$ curves (Fig. S2). The set-point oscillation amplitude of 100~pm was controlled to better than 2\% for the duration of the $z$ ramp (Fig. S1), and, in line with Peronio and Giessibl's measurement strategy~\cite{Peronio2016}, a sample voltage of $70~ \mathrm{\mu V}$ (compensating the overall offset bias in the control electronics) was used to minimize the tunneling current. The $\Delta f(z)$ curves were repeated over multiple individual measurements to both improve accuracy and, as described in detail below, derive statistical uncertainties; the average of all $B=0$~T~(blue) and $B=$~200~mT~(red) curves are shown in Fig.~2(a). The inset shows a similar averaged $\Delta f(z)$ curve, acquired over a larger $z$-range. During the frequency-shift measurements, the presence of the superconducting phase was routinely confirmed via scanning tunneling spectroscopy.
\flushbottom

To place a rigorous bound on any magnetic-field-induced change in the $\Delta f(z)$ curves, we first calculate the mean and variance for the sets of ``field'' ($B$) and ``no field'' ($B=0$) measurements: $\langle \Delta f\rangle_{\mathrm{B}}(z)$, $\sigma^2_{\mathrm{B}}(z)$, $\langle \Delta f\rangle_{\mathrm{B=0}}(z)$, and $\sigma^2_{\mathrm{B=0}}(z)$, where $\langle \rangle$ denotes the mean, $\sigma^2$ the variance. The combined standard error of the difference between these two sets is then
\begin{equation}
\mathcal{E}(z)=\sqrt{\frac{\sigma^2_{\mathrm{B}}(z)}{n_{B}} + \frac{\sigma^2_{\mathrm{B=0}}(z)}{n_{B=0}}},
\end{equation}
where $\sigma^2$ is the variance and $n_{\mathrm{B}}, n_{\mathrm{B=0}}$ are the number of $\Delta f(z)$ curves in the field and no-field sets, respectively. We calculate the difference between the two mean curves, $\langle \Delta f\rangle_{\mathrm{B=0}}(z) - \langle \Delta f\rangle_{\mathrm{B}}(z)$ (black line, Fig.~2b), and compare to the 99\% confidence interval (CI) derived from $\mathcal{E}(z)$ (shaded region, Fig.~2b). The measured difference lies predominantly within this CI, indicating that it is consistently below the threshold of experimental uncertainty. A root-mean-square (RMS) value for the difference over the entire curve in Fig.~2b, $D_{\mathrm{rms}} = 0.11$~Hz, is likewise well below the total RMS uncertainty of $0.17$~Hz. Our results therefore indicate, at a 99\% confidence level, that any superconductivity-induced change in the measured force-gradient curve must be smaller than 0.17~Hz RMS over the investigated distance range. If we use the simple dipole-dipole interaction model (Fig. S3) suggested by Peronio and Giessibl~\cite{Peronio2016}, this result constrains the upper limit on the change in the image dipole to 0.01~D. 

As an independent measurement of possible variation in electrostatic screening at the tip-sample junction, we turn to field emission resonance (FER) spectroscopy~\cite{Binnig1985, Becker1985, Ploigt2007}. FERs arise when electrons tunnelling from the tip become transiently trapped in quasi-bound states within the potential well bounded by the sample surface barrier and the tip~\cite{Becker1985,Binnig1985}. This well is formed by the combination of the image potential at the sample surface and the applied electric field; the former contributes significantly at low energies, while at sufficiently high positive sample bias the latter dominates, giving rise to a discrete ladder of quasi-triangular-well states -- essentially Gundlach oscillations~\cite{Lin2007} -- observed as a series of resonances in $\mathrm{d}I/\mathrm{d}V$ (or $\text{d}z/\text{d}V$) spectra in the field-emission regime~\cite{Binnig1985, Lin2007,Ploigt2007, Aladyshkin2020}. As such, the energies of the FER states are highly sensitive to the local electrostatic environment and have accordingly been used to map local work function variations with high spatial resolution~\cite{Ploigt2007, Lin2007, Aladyshkin2020}. Field emission resonance spectroscopy is therefore particularly well suited to testing whether electrostatic screening changes between the superconducting and normal states of Pb(111).

\begin{figure}
\includegraphics[width=1.02\linewidth]{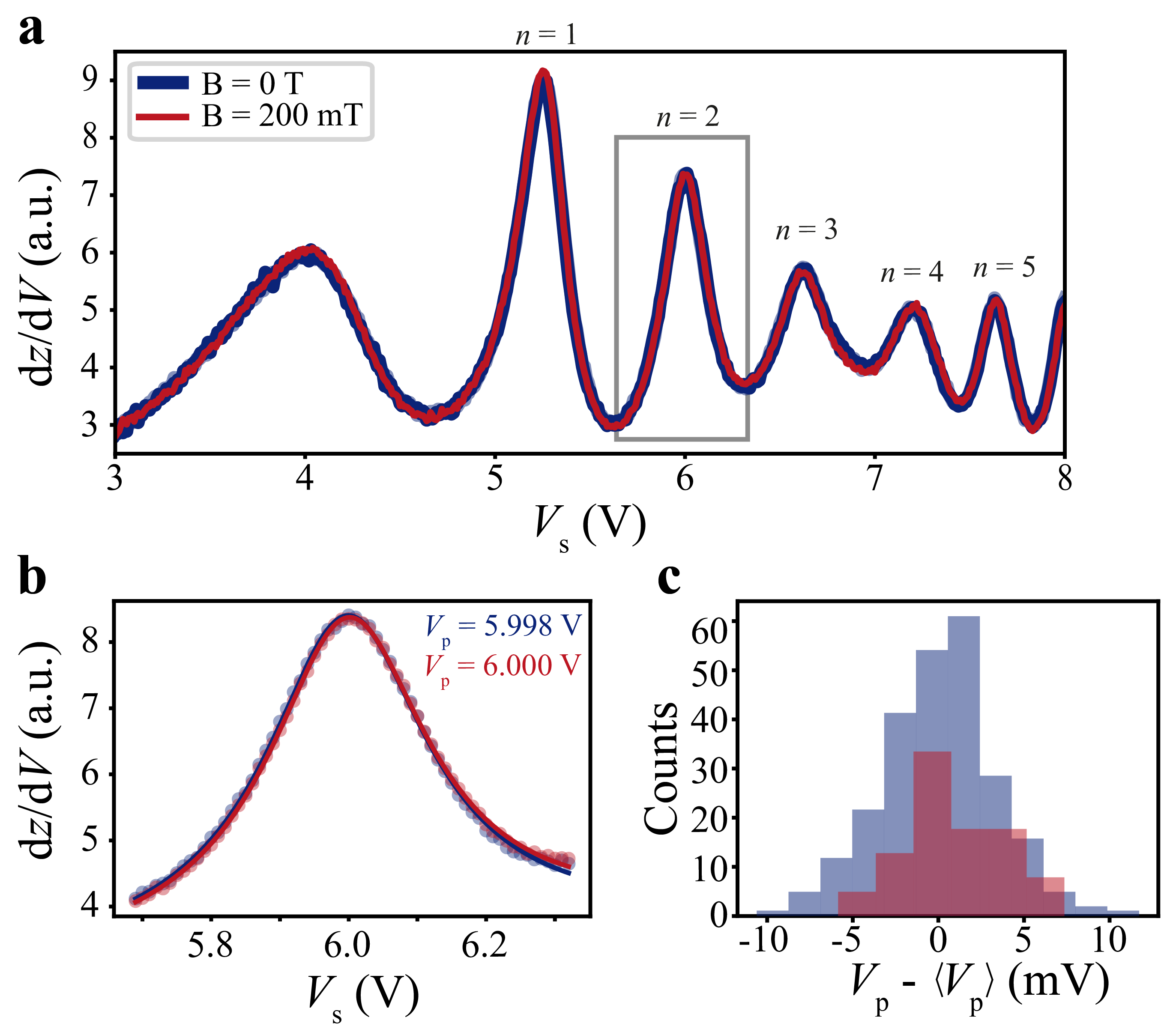}
\caption{(a) Constant-current $\text{d}z/\text{d}V$ spectrum as a function of applied bias in the field emission regime for no magnetic field (blue) and an applied $B$ field of 200 mT normal to the Pb(111) surface; (b) $n=2$ field emission resonance for both the zero-field (blue) and applied field (red) conditions, with Lorentzian fits (solid lines); (c) Histogram of the deviation of each measured FER peak position from its resonance-specific mean, pooled over $n=1,2,3$ and separated by field (red) and no-field (blue) conditions.}
\label{Fig3}
\end{figure}

FER spectra (Fig.~3) were acquired in constant-current mode~\cite{Binnig1985}, with $\text{d}I/\text{d}V$ spectra of the superconducting gap, alongside $\Delta f(z)$ and $I(z)$ curves, checked frequently between measurements to ensure that the tip state was preserved throughout. Fig. 3(a) shows a set of resonances for sample biases in the +3 V to +8 V range in the superconducting (blue) and normal (red) states. The FER spectra lie atop one another with no observable differences at this scale, including measurements made with varying tip apexes (Fig. S6). To establish a quantitative bound on this result, we restrict our analysis to the $n = 1,2,3$ FERs. Each individual resonance was fit with a Lorentzian peak to extract its peak center; an example for the $n=2$ resonance is shown in Fig. 3(b). Pure Lorentzian fits were robust over a range of $\pm 200$~mV around each peak center, $V_{\mathrm{p}}$, despite some asymmetry related to the spectral background often seen in tunneling spectroscopy. Repeated measurements under field and no-field conditions are shown in Fig.~3(c). For each resonance, we compare the peak position of an individual measurement, $V_{p,n}$, to the group average, $\langle V_{p,n}\rangle$; pooling this deviation across $n=1,2,3$ yields the histograms shown. The field (red) and no-field (blue) distributions entirely overlap, each with a mean difference of 0.0~mV and a 99\% confidence interval of $\pm0.5$~meV. The FER results are therefore entirely consistent with the force-gradient measurements of Fig. 2, confirming that the electrostatic response of the sample is indistinguishable between the superconducting and normal states. Any superconductivity-induced shift in the FER peak positions must be smaller than our experimental bound of 0.5~meV. We next convert this limit into a constraint on the magnitude of the image dipole.

\begin{figure}
\includegraphics[width=.98\linewidth]{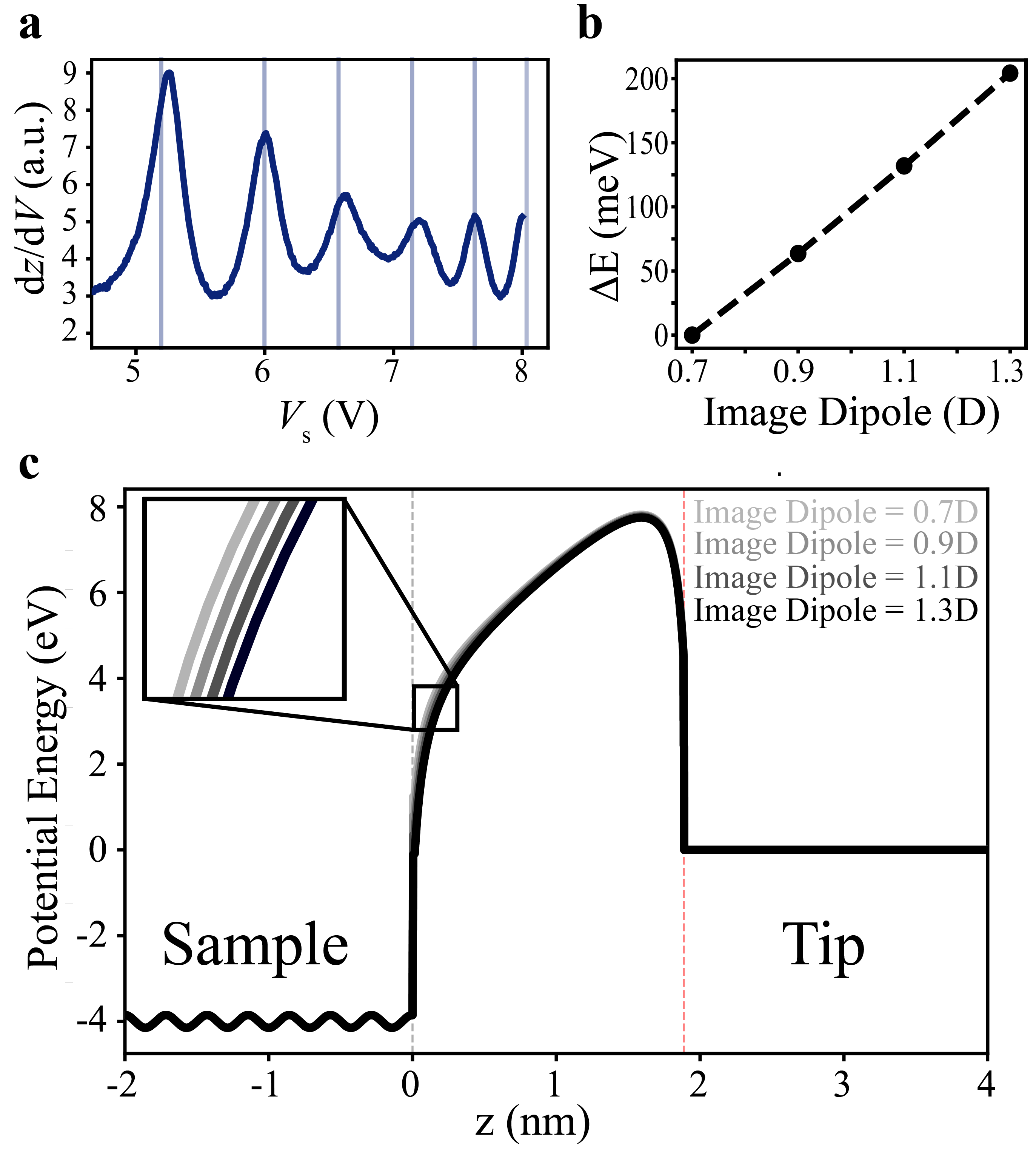}
\caption{(a) Comparison of experimentally measured peak positions of field emission resonances (blue line, no magnetic field) with those predicted by the simple model described in the text (vertical light blue lines, see also Supplementary Material); (b) predicted shift in resonance energy as a function of the magnitude of the image dipole; (c) electrostatic potential used in the simulations; the inset shows the influence of the image dipole on the form of the potential.}
\label{Fig4}
\end{figure}

In order to relate the 0.5 meV bound on shifts in the FER resonances to the local electrostatic response of the tip-sample junction, we constructed a model one-dimensional potential and solved the Schrödinger equation to find the characteristic energy eigenstates inside the vacuum region (Fig. 4), following the approach of Ploigt~\textit{et~al.}~\cite{Ploigt2007}. The potential used in the simulations is shown in Fig.~4(c): the Pb(111) crystal is modeled with a small periodic potential commensurate with the Pb lattice constant. A small number of physically motivated parameters -- the tip and image dipole moments, together with the sample work function and tip-sample separation -- are sufficient to account for the spectrum of observed FER energies (light blue lines, Fig.~4(a)), giving us confidence in the model's ability to adequately capture the underlying electrostatics. As a further test of the model's sensitivity to changes in the electrostatic energy landscape, we compared FER spectra measured on the Pb(111) terrace and at a step edge, the latter known to host a Smoluchowski dipole~\cite{Smoluchowski1941, Yu2006}; the model reproduces the difference between the two spectra well, accounting for it with only a small change in the sample dipole moment (see Fig. S5). We then translate the 0.5 meV bound on the resonance energy shift into a constraint on the image dipole: tuning the dipole moment and tracking the resulting shift in the simulated FER peak energies (Fig.~4(b)) gives a sensitivity of approximately 350~mV/Debye. This in turn constrains any superconductivity-induced change in the image dipole to less than 0.001~D -- an order of magnitude tighter than the 0.01~D bound obtained from the force-based $\Delta f(z)$ measurements.

In summary, the techniques employed here probe the electrostatic response of the superconducting condensate through entirely different physical channels -- one via the force gradient for a scanning probe, the other via the energy of quasi-bound electronic states in the tunnel junction -- yet arrive at the same conclusion: the image dipole at the tip-sample junction is unaffected, within tight experimental bounds (99\% confidence intervals), by the onset of superconductivity. Our results reaffirm that the Meissner effect is a purely transverse phenomenon, with no accompanying change in the superconductor's response to static electric fields. More broadly, this work demonstrates the sensitivity that combined force- and spectroscopy-based scanning probe techniques bring to the electrostatics of the superconducting state, an approach that could be extended to electrostatically gated systems and unconventional superconductors~\cite{Cao2018, Oh2021}.

\textit{Acknowledgments} We gratefully acknowledge technical assistance from Nick Botterill and Yutaka Miyatake. This work was supported by the Engineering and Physical Sciences Research Council, grant reference EP/Y023250/1. PM thanks the EPSRC for the award of an Established Career Fellowship (EP/T033568/1).

\textit{Data Availability} All raw data from this work are available at the University of Nottingham Research Data Management Repository (https://doi.org/10.17639/nott.39715).

\bibliography{HirschBib}


\newpage

\setcounter{figure}{0}
\renewcommand{\thefootnote}{\fnsymbol{footnote}}
\makeatletter 
\renewcommand{\thefigure}{S\@arabic\c@figure}
\makeatother

\begin{center}
\textbf{\large SUPPLEMENTARY MATERIAL}
\end{center}

\section{Methods} 

 Our scanning tunneling microscopy/spectroscopy and qPlus atomic force microscopy measurements were carried out using a Unisoku USM-1300 system with a base pressure of $\sim 1 \times 10^{-10}$ mbar. The microscope is cooled to 4.2 K via a surrounding $^4$He bath (whose degree of thermal coupling to the system can be controlled through the introduction/removal of helium gas to/from an intervening vacuum space) and then to sub-400 mK temperatures using a combination of cooling via a 1K-pot and subsequent $^3$He condensation/evaporation. The results discussed in this paper are all acquired between 330-360~mK. The temperatures we report are the values recorded by a combination of a Cernox$^{\mathrm{TM}}$ sensor and a Lakeshore controller.

 The Pb(111) single crystal (MaTecK GmbH, J\"ulich) was prepared by repeated sputter-annealing cycles (500~eV~Ar$^+$, $\sim 1 \times 10^{-5}$ mbar) until flat terraces, of the order of tens of nanometers wide, with a low defect density were observed in STM overview images. The Pb(111) crystal was radiatively heated using a filament on the sample plate. We infer an upper bound of $\sim$~450~K, although the sample temperature during annealing was not measured directly. The Pb(111) step height of 283~pm (at 363~mK) provided a natural calibration standard for the z-axis of the piezoelectric transducer. 

 Nanonis Mimea$^{\mathrm{TM}}$ electronics and software were used to control, monitor, and record all data acquisition. A FEMTO Messtechnik GmbH DLPCA-200 transimpedance amplifier was used for all tunnel current measurements. For dI/dV measurements of the superconducting gap a modulation amplitude of 20 $\mu$V was applied to the tip-sample bias (via a 1/100 voltage divider to attenuate the output from the digital-to-analog converter (DAC)) and the Nanonis lock-in amplifier module was used to demodulate the tunnel current signal. Image potential/field emission resonance spectra were acquired via $dI/dV$ spectroscopy taken at constant-current (denoted $dI/dV_I$) with a lock-in modulation of 20 mV.   

The qPlus tuning fork sensor had a resonance frequency of 28.763~kHz and a quality factor (at 363~mK) of $\sim$ 38,000. Unless otherwise noted, the oscillation amplitude of the tuning fork ($A$) was 100~pm. 
 Given the central importance of accurate tip positioning/displacement especially in the $z$ axis, i.e. normal to the surface, for the rigorous comparison of measurements made with and without an applied magnetic field, we routinely monitored piezo drift/creep (for all three axes) throughout the experiment. Below 400~mK, drift for the $x,y$, and $z$ axes was $\sim$ 11~fm/s, $\sim$ 4~fm/s, and $\sim$ 8~fm/s, respectively, measured over the course of 10~hours ($x,y$) or 1.6~hours ($z$). Moreover, in all cases we acquired $\Delta f(z),I(z),$ and $dI/dV_I$ spectra in both the ``forward'' and ``backward'' directions to monitor any influence of hysteresis of the piezoelectric actuators and/or to check for the onset of non-conservative forces (dissipation).  

\begin{figure}[b!]
\includegraphics[width=1\linewidth]{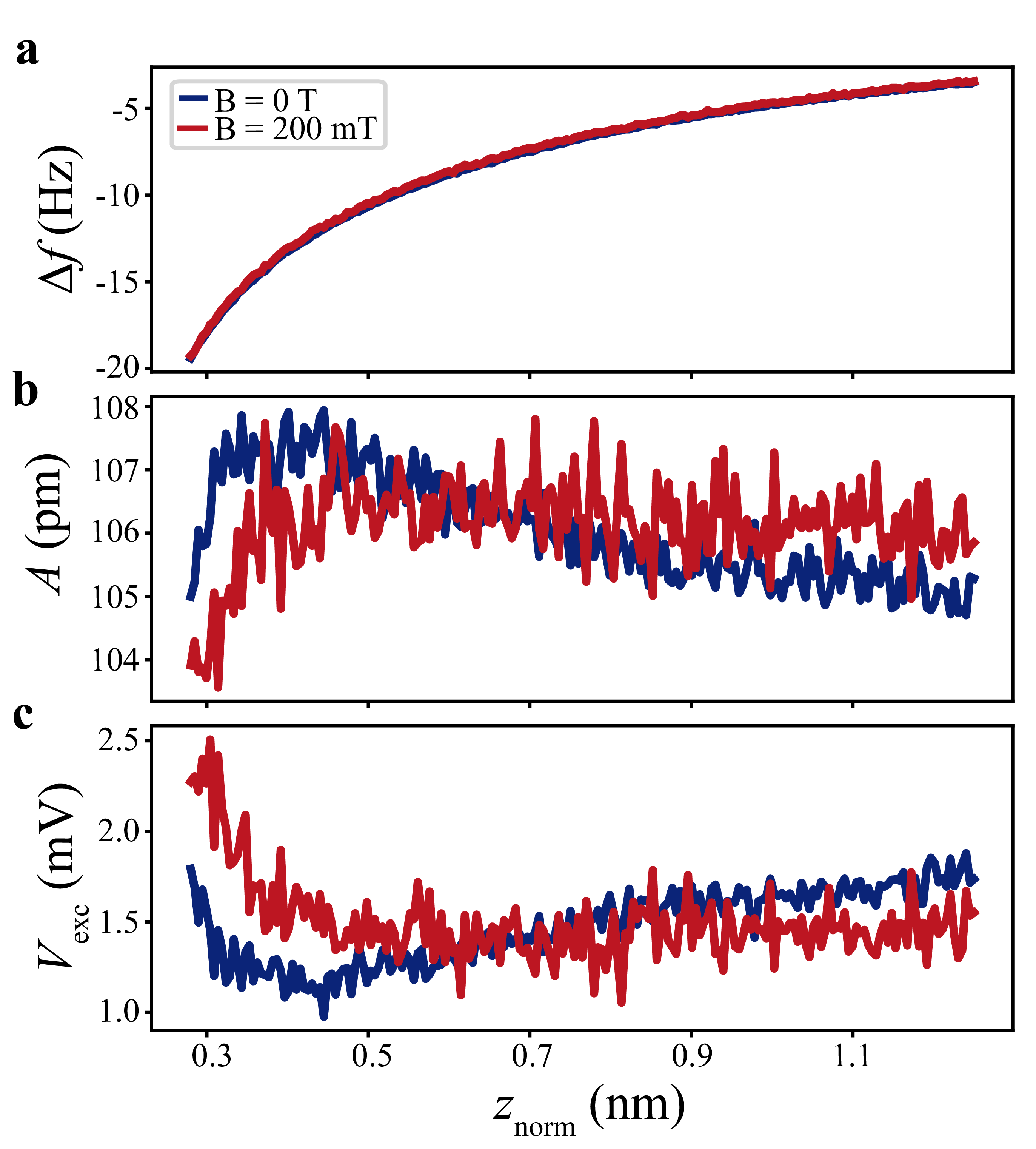}
\caption{(a) Frequency shift measured as a function of normalized tip-sample separation ($z_{norm}$). (b) Oscillation amplitude and (c) excitation voltage over the same $z_{norm}$ range}.
\label{Fig1}
\end{figure}

\section{$\Delta f(z)$ measurement protocol} 
We measure $\Delta f (z)$ curves by ramping $z$ through two primary ranges with different $z$ sampling resolutions:~(i)~\mbox{$ 0.5~\text{nm} < z_{norm} < 4.5~\text{nm}$} and (ii)~\mbox{$ 0.28~\text{nm} < z_{norm} < 1.25~\text{nm}$}, where $z_{norm}$ is defined in Fig. S2 below. Multiple intervals were included to ensure that the tip-sample interactions converged to the same behavior in the long-range limit, while simultaneously probing close enough to observe small changes in the short-range forces. Individual $\Delta f (z)$ curves were repeated between 10 and 60 times, with the $\Delta f$, amplitude, excitation, and phase channels collected in parallel in both the forward and backwards direction (see Fig. S1). Acquiring measurements over multiple $\Delta f(z)$ ranges in this fashion not only improved the statistical robustness of the analysis by enabling cross-comparison between different displacement intervals, but also meant we could minimize the effects of creep and drift while also probing as wide a spatial range of the interaction potential as possible. The closest approach to the sample surface, $z_{norm} = 280~\text{pm}$, was determined by carefully monitoring the excitation signal during the approach and ensuring that there was no hysteresis in any of the channels when measuring in the forward/backward direction. Any systematic deviation of the excitation signal from its value far from the surface was taken as an indication of mechanical relaxations, and the measurements were rejected (Fig. S1).

\section{Calculating absolute tip-sample distance} 

To compare the measured frequency-shift-versus-distance curves with the range predicted by the dipole-dipole interaction model, the absolute tip-sample separation was estimated. We employed the conventional approximation of defining $z=0$ at the point-contact position, corresponding to a tunneling conductance of \[ G_0 = \frac{2e^2}{h} \approx 77.5\,\mu\mathrm{S}, \] for a non-oscillating sensor. The point-contact position was obtained by linearly fitting the log of the tunneling current as a function of tip-sample distance and extrapolating to $G_0$ (Fig. S2). This offset was subsequently used to calculate the normalized $z$ position applied to all datasets to place the distance axis on an approximate absolute scale.

\begin{figure}[b!]
\includegraphics[width=1.0\linewidth]{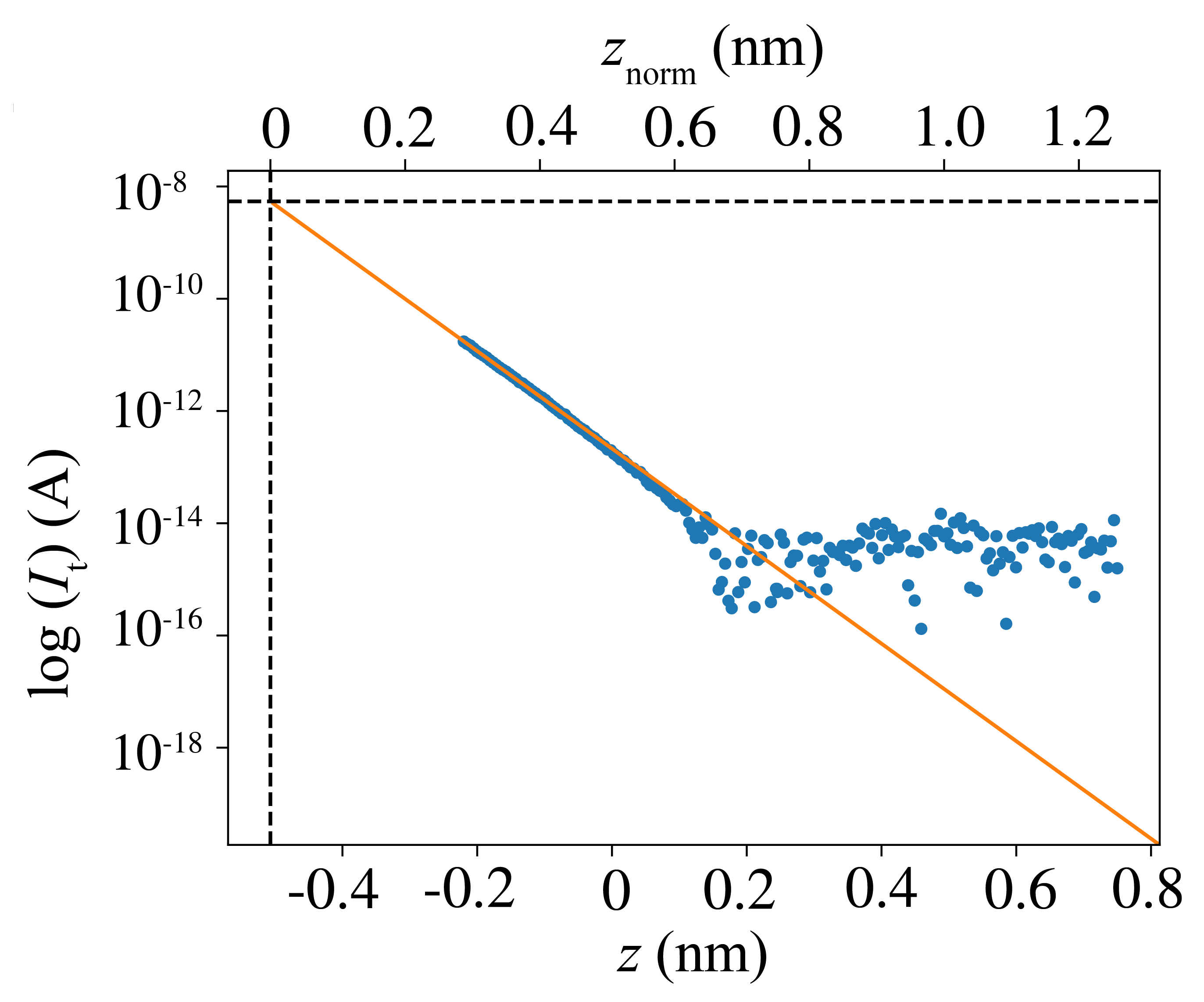}
\caption{Tunneling current measured as a function of the relative tip-sample distance (bottom y-axis) and absolute tip-sample distance (top y-axis) plotted in log scale (blue circles). A linear fit (orange) to the exponential decay has been fit for a distance range between $z_{norm}=0.28~\text{nm}$ and $z_{norm}~=~0.50~\text{nm}$.} 
\label{Fig2}
\end{figure}

\section{Dipole-dipole interaction} 

To estimate the maximum change in frequency shift due to changes in dipole screening, we employ a simple dipole-dipole interaction model~[S1]. The model consists of two dipoles separated by a distance $z$, as illustrated in the inset of Fig.~S3. The interaction force is calculated as a function of $z$ and subsequently converted into a frequency shift. 
\begin{equation} F_z = -\frac{12}{4\pi\epsilon_0}\frac{p^2}{(2z)^4}. \label{eq:Fz} \end{equation}
\begin{equation} \Delta f = -\frac{f_0}{2k}\left\langle \partial_z F_z \right\rangle = -\frac{f_0}{2k}\, \frac{3}{4\pi\epsilon_0}\, p^2 \left\langle z^{-5}\right\rangle . \label{eq:df}
\end{equation}

\begin{figure}[b!]
\includegraphics[width=1.02\linewidth]{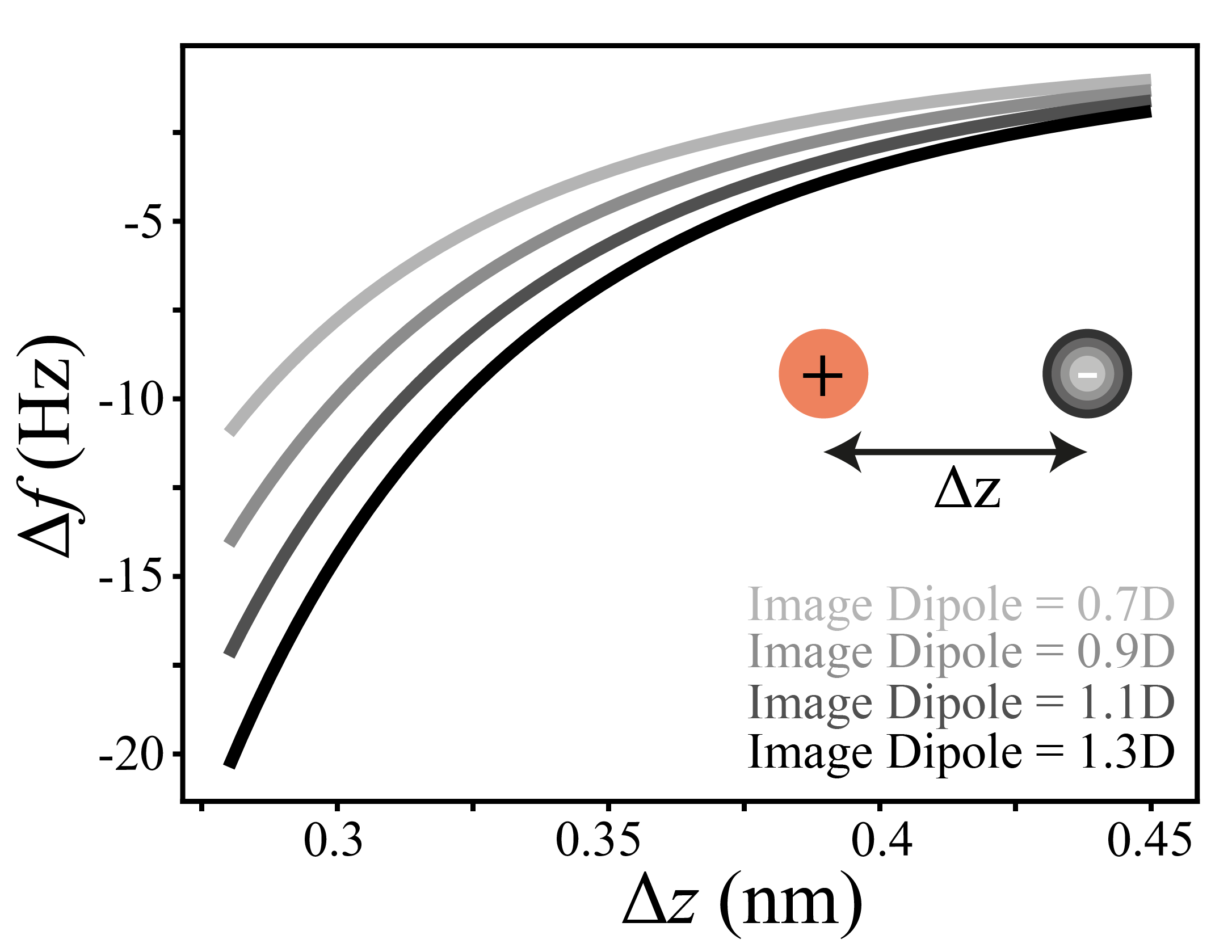}
\caption{Frequency shift as a function of tip-sample distance between two dipoles (depicted for simplicity as orange and black point charges). Different shades of gray depict the varying magnitude of the image dipoles (shades of gray) at distances calculated based on the absolute distance observed in experiments.}
\label{Fig3}
\end{figure}

\begin{figure}[t!]
\includegraphics[width=1.0\linewidth]{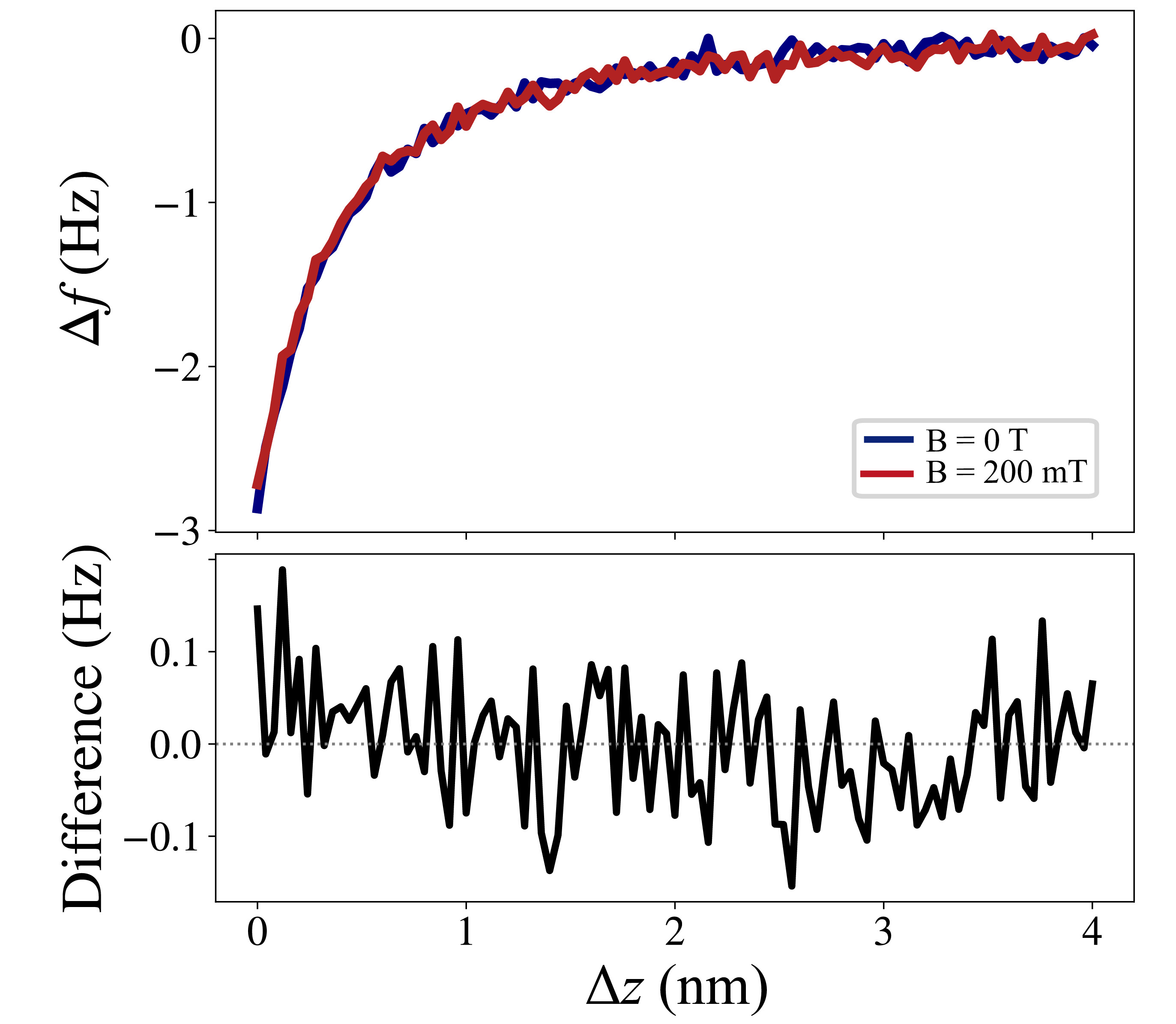}
\caption{(a) Averaged set of $\Delta f(z)$ measurements on the Ag(111) surface with $B=0$~T~(blue) and $B=200$~mT~(red), respectively. (b) Difference signal between $B~=~0$~T and $B~=~200$~mT.  }
\label{Fig4}
\end{figure}

Fig.~S3 shows the resulting frequency shift as a function of $z$ as one of the dipole moments is modified in magnitude. This dipole variation is intended to mimic changes in the image dipole strength which would arise from modifications to the electrostatic screening. The expected frequency shifts are relatively large (several Hz) given small changes (0.2 D) in the "image" dipole moment in this calculation. These expected changes will later be used to provide a quantitative upper limit on the expected change in the image dipole when transitioning between the superconducting and normal states. We note, however, that this simple model takes no account of background, integrated van der Waals interactions.

\section{Control measurements on a normal metal} 

Fig.~S4 presents a control measurement of the frequency shift as a function of tip-sample distance acquired above Ag(111), a non-superconducting, normal metal at magnetic fields of 0~T (blue) and 200~mT (red), where 200~mT is the field used to quench superconductivity in Pb(111). The lower panel displays the point by point difference between the two datasets. Within the experimental resolution, no magnetic field-induced change is detected, with all observed deviations remaining below the previously determined uncertainty of 0.17~Hz reported in the main text. This measurement further demonstrates the limited impact of the magnetic field on the sensitivity of the force-based measurement in this work. 

\section{Identifying dipoles at step edges using FER} 

Metal step edges are known to exhibit Smoluchowski dipoles due to the abrupt nature of the physical system. Constant-current $\mathrm{d}z/\mathrm{d}V$ spectroscopy at high bias voltages reveals field-emission resonances (FERs), whose energies are sensitive to the local electrostatic potential and, consequently, to the presence of these dipoles. Fig.~S5 compares FER spectra acquired on a Pb(111) terrace (blue) and at an atomic step edge (green). A pronounced shift in the resonance energies is observed at the step edge. The vertical solid lines denote the eigenenergies obtained from the one-dimensional image-potential model described in the main text~[S2], which reproduces the measured resonance positions well. Parameterizing the spectra within this model yields a dipole-moment difference of approximately $0.2~\mathrm{D}$ at the step edge, in good agreement with previous observations of Smoluchowski dipole moments at metallic step edges~[S3,S4]. The success of the model in capturing the expected dipole at the atomic step edge further validates its use in the quantification of the screening measurement in the main text. 

\begin{figure}[t!]
\includegraphics[width=1.0\linewidth]{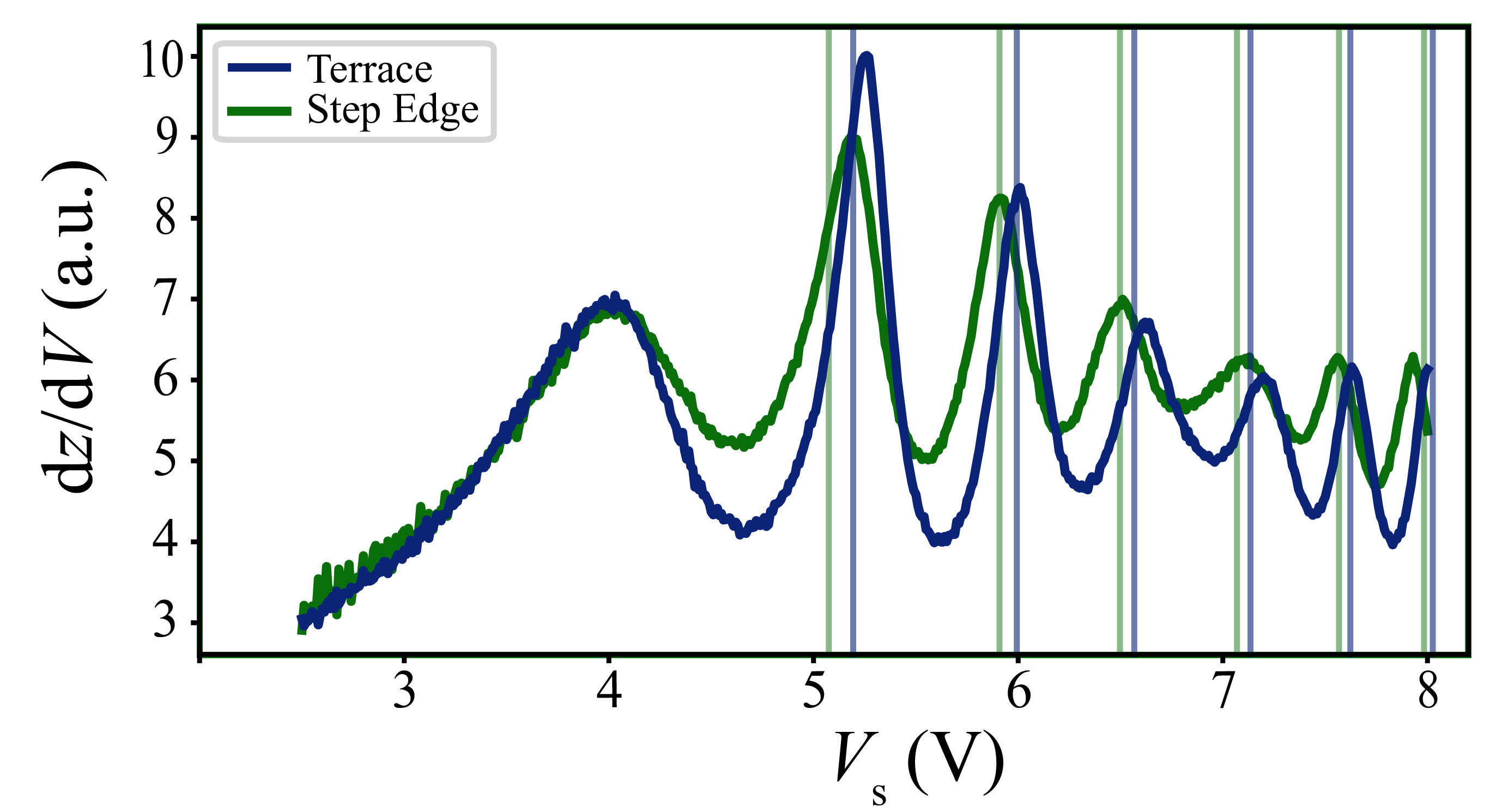}
\caption{a) Constant-current ($\text{d}z/\text{d}V$) spectrum as a function of $V_S$ in the field emission regime on a Pb(111) terrace (blue) and a step edge on Pb(111).}
\label{Fig5}
\end{figure}

\section{Field emission resonances with different tips} 

\begin{figure}[t!]
\includegraphics[width=.95\linewidth]{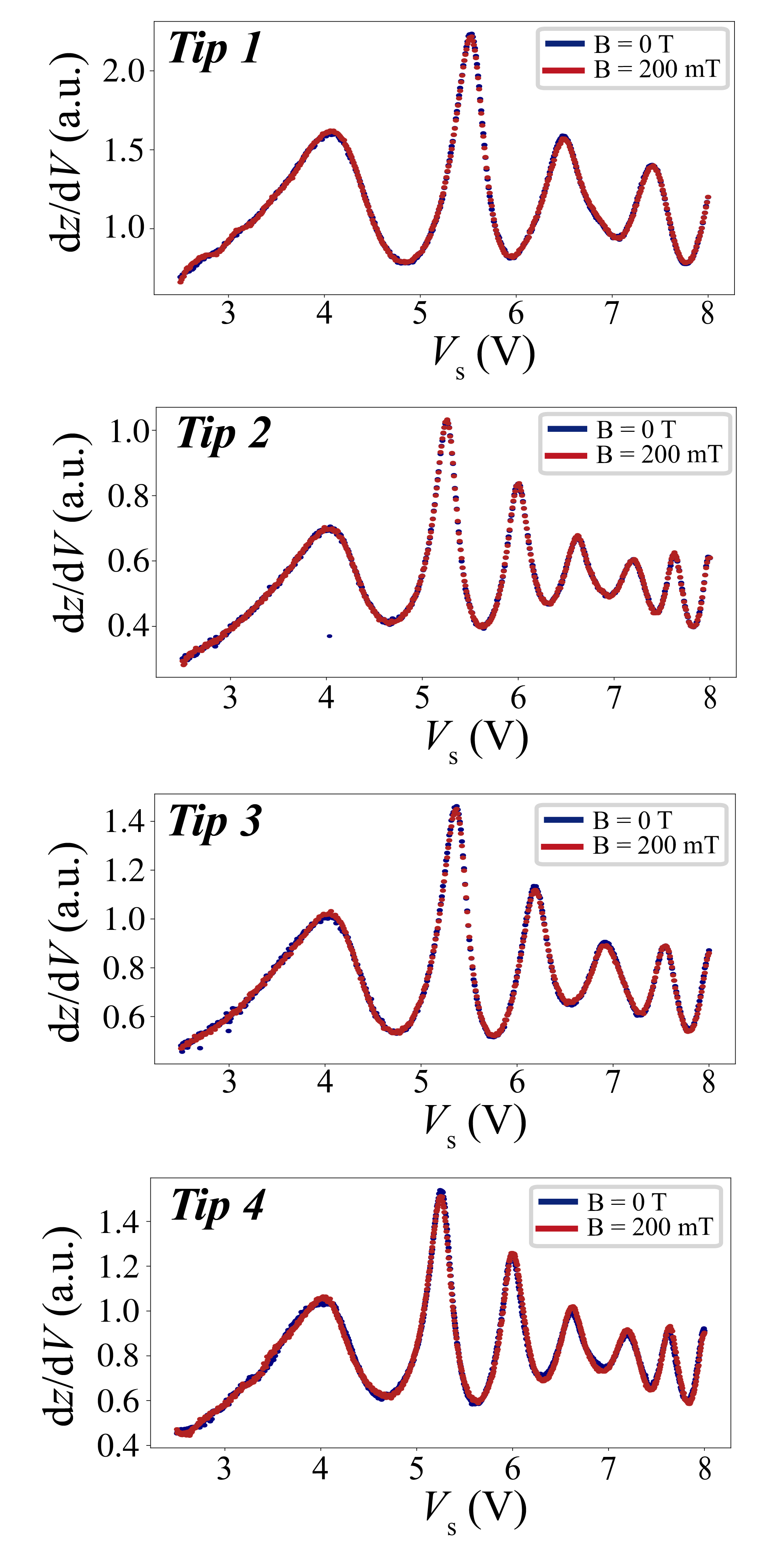}
\caption{Constant-current ($\text{d}z/\text{d}V$) spectrum as a function of applied bias in the field emission regime for no magnetic field (blue) and an applied $B$ field of 200 mT normal to the Pb(111) surface (red) with a different tip apex in each case.}
\label{Fig6}
\end{figure}

Fig.~S6 presents constant-current ($\mathrm{d}z/\mathrm{d}V$) spectra acquired on a Pb(111) terrace, analogous to the measurements discussed in Fig.~3 of the main text, using four distinct tip apexes. For each tip, spectra were acquired at 0~T (blue), corresponding to the superconducting state, and at 200~mT (red), where superconductivity is quenched. In all cases, the spectra recorded at the two magnetic fields are virtually indistinguishable, with the resonance energies and spectral line shapes coinciding within the experimental uncertainty. The consistency of this result across multiple tip terminations rules out tip-specific effects and confirms that no detectable change in the FER spectrum occurs upon crossing the superconducting phase transition. This robustness further supports the conclusion that any superconductivity-induced modification of the local electrostatic screening is below the sensitivity limit of our measurements.

\begin{center}
\textbf{Supplementary References}
\end{center}
\small
\begin{enumerate}
\item[{[S1]}] A. Peronio and F. J. Giessibl, \textit{Phys.~Rev.~B}~\textbf{94},~094503~(2016).
\item[{[S2]}] H.-C. Ploigt, C. Brun, M. Pivetta, F. Patthey, and W.-D. Schneider, \textit{Phys.~Rev.~B}~\textbf{76},~195404~(2007).
\item[{[S3]}] J. F. Jia, K. Inoue, Y. Hasegawa, W. S. Yang, and T. Sakurai, \textit{Phys.~Rev.~B}~\textbf{58},~1193~(1998).
\item[{[S4]}] J. Y. Park, G. M. Sacha, M. Enachescu, D. F. Ogletree, R. A. Ribeiro, P. C. Canfield, C. J. Jenks, P. A. Thiel, J. J. S\'aenz, and M. Salmeron, \textit{Phys.~Rev.~Lett.}~\textbf{95},~136802~(2005).
\end{enumerate}

\end{document}